\documentclass[10pt,letterpaper]{article}

\usepackage{cogsci}
\usepackage{pslatex}
\usepackage{apacite}
\usepackage{graphicx}

\title{Society Functions Best with an Intermediate Level of Creativity}
 
\author{{\large \bf Liane Gabora (liane.gabora@ubc.ca)}\\
  University of British Columbia (Okanagan campus)\\
  Department of Psychology, Arts Building, 3333 University Way\\
  Kelowna BC, V1V 1V7, CANADA\\
  \AND {\large \bf Hadi Firouzi (hadi.firouzi@ubc.ca)}\\
  University of British Columbia (Okanagan campus)\\
  Department of Engineering, EME Building, 3333 University Way\\
  Kelowna BC, V1V 1V7, CANADA\\
}

\begin{document}
\maketitle
\noindent {Reference: Gabora, L., \& Firouzi, H. (2012). Society functions best with an intermediate level of creativity. Proceedings of the Annual Meeting of the Cognitive Science Society (pp. 1578-1583). August 1-4, Sapporo Japan. Houston TX: Cognitive Science Society.}

\begin{abstract}
In a society, a proportion of the individuals can benefit from creativity without being creative themselves by copying the creators. This paper uses an agent-based model of cultural evolution to investigate how society is affected by different levels of individual creativity. We performed a time series analysis of the mean fitness of ideas across the artificial society varying both the percentage of creators, $C$, and how creative they are, $p$ using two discounting methods. Both analyses revealed a valley in the adaptive landscape, indicating a tradeoff between $C$ and $p$. The results suggest that excess creativity at the individual level can be detrimental at the level of the society because creators invest in unproven ideas at the expense of propagating proven ideas.\\

\end{abstract}
 
\section{Introduction}

 \noindent 
Our capacity for self-expression, problem solving, and making aesthetically pleasing artifacts, all stem from our creative abilities. Psychologists have almost universally converged on the definition of creativity proposed by Guilford over sixty years ago at his annual address to the American Psychological Association \cite{Moran2011}. Guilford defined creativity in terms of two criteria: originality or novelty, and appropriateness or adaptiveness, \emph{i.e.,} relevance to the task at hand. 
Individuals vary from not particularly creative to highly creative. Not only are humans individually creative, but we build on each other's ideas such that over centuries, art, science, and technology, as well as customs and folk knowledge, can be said to evolve \cite{Cavalli-SforzaFeldman1981,Gabora1996,MesoudiWhitenLaland2006,WhitenHindeLalandStringer2011}. Creativity has long been associated with personal fulfillment \cite{May1975,Rogers1959}, self-actualization \cite{Maslow1959}, and maintaining a competitive edge in the marketplace, and it is often assumed that more creativity is necessarily better. 

However, there are significant drawbacks to creativity \cite{CropleyCropleyKaufmanRunco2010,Ludwig1995}. Generating creative ideas is time consuming, and a creative solution to one problem often generates other problems, or has unexpected negative side effects that may only become apparent after much effort has been invested. Creative people often reinvent the wheel, and may be more likely to bend rules, break laws, and provoke social unrest \cite{SternbergLubart1995,Sulloway1996}. They 
are more prone to affective disorders such as depression and bipolar disorder, and have a higher incidence of schizophrenic tendencies, than other segments of the population \cite{Andreason1987,Flaherty2005,GoodwinJamieson1990}. 

Given these negative aspects of creativity, it is perhaps just as well that, in a group of interacting individuals, not all of them need be particularly creative for the benefits of creativity to be felt throughout a social group. The rest can reap the rewards of the creator's ideas by simply copying, using, or admiring them. Few of us know how to build a computer, or write a symphony, but they are nonetheless ours to use and enjoy. Clearly if everyone relied on the strategy of imitating others, the generation of cultural novelty would grind to a halt. This raises the following questions: what is the ideal ratio of creators to imitators, and how creative should the ``creative types'' be? 

\section{The Model}

We investigated this using an agent-based model of cultural evolution referred to as ``EVOlution of Culture'', abbreviated EVOC. To our knowledge, EVOC is the only computational model that enables one to create an artificial world with agents of varying levels of creativity and observe the effect of varying creativity on mean fitness and diversity of ideas in the artificial society. It uses neural network based agents that (1) invent new ideas, (2) imitate actions implemented by neighbors, (3) evaluate ideas, and (4) implement successful ideas as actions. EVOC is an elaboration of Meme and Variations, or MAV \cite{Gabora1995}, the earliest computer program to model culture as an evolutionary process in its own right, as opposed to modelling the interplay of cultural and biological evolution\footnote{The approach can thus be contrasted with computer models of how individual learning affects biological evolution \cite{Best1999,Best2006,Higgs2000,HintonNowlan1987,HutchinsHazelhurst1991}.}. The approach was inspired by genetic algorithm \cite{Holland1975}, or GA. The GA is a search technique that finds solutions to complex problems by generating a ÔpopulationÕ of candidate solutions through processes akin to mutation and recombination, selecting the best, and repeating until a satisfactory solution is found. The goal behind MAV, and also behind EVOC, was to distil the underlying logic of not biological evolution but cultural evolution, \emph{i.e.,} the process by which ideas adapt and build on one another in the minds of interacting individuals. Agents do not evolve in a biological sense--they neither die nor have offspring--but do in a cultural sense, by generating and sharing ideas for actions. In cultural evolution, the generation of novelty takes place through invention instead of through mutation and recombination as in biological evolution, and the differential replication of novelty takes place through imitation, instead of through reproduction with inheritance as in biological evolution. EVOC has been used to address such questions as how does the presence of leaders or barriers to the diffusion of ideas affect cultural evolution.

We now summarize briefly the architecture of EVOC in sufficient detail to explain our results; for further details on the model, we refer the reader to previous publications \cite{Gabora1995,Gabora2008a,Gabora2008b,GaboraLeijnen2009,LeijnenGabora2009,GaboraSaberi2011}.

\subsection{Agents}
Agents consist of (1) a neural network, which encodes ideas for actions and detects trends in what constitutes a fit action, (2) a `perceptual system', which carries out the evaluation and imitation of neighbours' actions, and (3) a body, consisting of six body parts which implement actions. The neural network is composed of six input nodes and six corresponding output nodes that represent concepts of body parts (LEFT ARM, RIGHT ARM, LEFT LEG, RIGHT LEG, HEAD, and HIPS), and seven hidden nodes that represent more abstract concepts (LEFT, RIGHT, ARM, LEG, SYMMETRY, OPPOSITE, and MOVEMENT). Input nodes and output nodes are connected to hidden nodes of which they are instances (\emph{e.g.,} RIGHT ARM is connected to RIGHT.) Each body part can occupy one of three possible positions: a neutral or default positions, and two other positions, which are referred to as active positions. Activation of any input node activates the MOVEMENT hidden node. Same-direction activation of symmetrical input nodes (\emph{e.g.,} positive activation--which represents upward motion--of both arms) activates the SYMMETRY node. In the experiments reported here the OPPOSITE hidden node was not used.

\subsection{Invention}
An idea for a new action is a pattern consisting of six elements that dictate the placement of the six body parts. Agents generate new actions by modifying their initial action or an action that has been invented previously or acquired through imitation. During invention, the pattern of activation on the output nodes is fed back to the input nodes, and invention is biased according to the activations of the SYMMETRY and MOVEMENT hidden nodes. (Were this not the case there would be no benefit to using a neural network.) To invent a new idea, for each node of the idea currently represented on the input layer of the neural network, the agent makes a probabilistic decision as to whether the position of that body part will change, and if it does, the direction of change is stochastically biased according to the learning rate. If the new idea has a higher fitness than the currently implemented idea, the agent learns and implements the action specified by that idea. 

\subsection{Imitation}
The process of finding a neighbour to imitate works through a form of lazy (non-greedy) search. The imitating agent randomly scans its neighbours, and adopts the first action that is fitter than the action it is currently implementing. If it does not find a neighbour that is executing a fitter action than its own current action, it continues to execute the current action. 

\subsection{Evaluation}
\indent Following \citeA{Holland1975}, we refer to the success of an action in the artificial world as its \emph{fitness}, with the caveat that unlike its usage in biology, here the term is unrelated to number of offspring (or ideas derived from a given idea). Fitness of an action is determined using a predefined equation{, Eq.}~\ref{eq:fitness}{,} that ascribes a range of fitness values from 0 to 10 to the 729 possible actions. (Six body parts that can be in three possible positions gives a total of 729.) The fitness function used in these experiments rewards activity of all body parts except for the head, and symmetrical limb movement. Total body movement, $m$, is calculated by adding the number of active body parts, \emph{i.e.,} body parts not in the neutral position. The fitness $F$ of an action is calculated as follows:

\begin{equation}
	F_{nc} = m + 1.5 (s_a+s_t)+2(1-m_h)
\label{eq:fitness}
\end{equation}
$s_a = 1$ if arms are moving symmetrically; 0 otherwise \\
$s_t = 1$ if legs are moving symmetrically; 0 otherwise \\
$m_h = 1$ if head is stationary; 0 otherwise \\

\noindent Note that actions have a cultural version of what in biology is referred to as epistasis, wherein what is optimal with respect to one component depends on what is done with respect to another. Epistasis occurs because what is optimal for the left arm depends on what the right arm is doing, and \emph{vice versa,} and same for the legs. 

\subsection{Learning}
Invention makes use of the ability to detect, learn, and respond adaptively to trends. Since no action acquired through imitation or invention is implemented unless it is fitter than the current action, new actions provide valuable information about what constitutes an effective idea. Knowledge acquired through the evaluation of actions is translated into educated guesses about what constitutes a successful action by updating the learning rate. For example, an agent may learn that more overall movement tends to be either beneficial (as with the fitness function used here) or detrimental, or that symmetrical movement tends to be either beneficial (as with the fitness function used here) or detrimental, and bias the generation of new actions accordingly.

\subsection{A Typical Run}
Fitness of actions starts out low because agents are initially immobile. They are all implementing the same action, with all body parts in the neutral position; thus action diversity is at a minimum. Soon some agent invents an action that has a higher fitness than immobility, and this action gets imitated, so fitness increases. Fitness increases further as other ideas get invented, assessed, implemented as actions, and spread through imitation. The diversity of actions increases due to the proliferation of new ideas, and then decreases as agents hone in on the fittest actions. In the version of the model used here, fitness values hit a ceiling and converge\footnote{This is not the case for another version of the model \cite{GaboraSaberi2011}.}. Thus, over successive rounds of invention and imitation, the agents' actions improve. EVOC thereby models how ``descent with modification'' occurs in a purely cultural context. 

\section{Experiments}
To carry out our investigation of how varying the level of creativity of individuals affects the fitness of ideas in society as a whole, these experiments used a default artificial world: a toroidal lattice with 1024 nodes, each occupied by a single, stationary agent, and a von Neumann neighborhood structure (agents only interacted with their four adjacent neighbors). Creators and imitators were randomly dispersed.\footnote{In other experiments \cite{LeijnenGabora2009} we investigated the results of clustering creators.} Runs lasted 100 iterations.

\indent In an earlier version of EVOC, in which the ratio of inventing and imitating was always the same for all agents, we found that the society as a whole did best when the ratio of creating to imitating was approximately 2:1 \cite{Gabora1995}. To incorporate individual differences in degree of creativity, we constructed a version of EVOC that enables us to distinguish two types of agents: \emph{imitators}, that only obtain new actions by imitating neighbors, and \emph{creators}, that obtain new actions by either  inventing one or by imitating a neighbor. Imitators never invent at all; they simply copy the creators' successful inventions. Thus all new actions are generated by creators. We also made it possible to vary the probability that creators create versus imitate; each agent can be a pure imitator, a pure creator, or something in between. Whereas any given agent is either a creator or an imitator throughout the entire run, the proportion of creators innovating or imitating in a given iteration fluctuates stochastically. The proportion of creators relative to imitators in the society is referred to as $C$. The creativity of the creators -- that is, the probability that a creator invents a new action instead of imitating a neighbor -- is referred to as $p$. If a creator decides to invent on a particular iteration, the probability of changing the position of any body part involved in an action is 1/6.\footnote{This gave on average a probability of one change per newly invented action, which previous experiments \cite{Gabora1995} showed to be optimal.} 

The society consists of three subgroups:

\begin{itemize}
\item {$C \times p \times N$} creators attempting to innovate
\item {$C \times (1-p) \times N$} creators attempting to imitate
\item {$(1-C) \times N$} imitators attempting to imitate
\end{itemize}
{\noindent where the number of agents, N is 1024.}

In previous investigations we measured, for different values of $C$ and $p$, the diversity of ideas over the course of a run. We found that the cultural diversity, \emph{i.e.,} the number of different ideas implemented by one or more agent(s), was positively correlated with both the proportion of creators to imitators, and with how creative the creators were. We also obtained suggestive evidence that when creators are relatively uncreative, the mean fitness of ideas increases as a function of the percentage of creators in the society, but when creators are highly creative, the society appears to be better off with fewer creators \cite{LeijnenGabora2009}. However, those simulations were performed with small societies (100 agents), and since action fitness was obtained at only one time slice (after 50 iterations) for all ratios of creators to inventors, these results did not reflect the dynamics of the time series. Given a set of series of accumulating value over time, it is unclear which series is most representative. The series cannot be unambiguously ordered unless for each pair of series one strictly dominates the other, and that is not the case here; the curves representing mean fitness at different values of $\{C, p\}$ increase monotonically but they often cross and re-cross as time progresses. Thus here we present a more extensive investigation of the relationship between creativity and society as a whole that employs a sophisticated solution to the time series problem.

\section*{Analysis}

\indent We used time series discounting which associates a ``present value'' with any future benefit such that the present value of any given benefit diminishes as a function of elapsed time until the benefit is realized \cite{McDonaldSiegel1986}. The standard approach in financial settings is exponential discounting. Given a series of benefits $b_{t}$, the Net Present Value (NPV) is defined as:

\begin{equation}
NPV(b) =  \displaystyle\sum_{t=1}^N r^{t-1}  b_{t} \quad with \quad 0 < r \leq 1
\label{eq:npv}
\end{equation}

\indent The discount rate $r$ is normally set as $r = (\frac{100+i}{100})^{-1}$ where $i$ is the interest rate (in percentage) for the unit period that an investor can obtain from a safe investment. 

This basic idea was adapted to analyze the benefit accrued by attaining fit actions for different values of $C$ and $p$ in EVOC. The first discounting method used was Time-to-Threshold (TTT) discounting. Since all fitness trajectories were monotonically increasing, those that reached a reasonably high threshold $\tau$ sooner should be valued higher. We measured how many iterations (time to threshold) it took for fitness to reach $\tau$. For these runs, $\tau = 9$ was used as a measure of optimal fitness to allow for a realistic averaging over time. 

\indent Whereas imitators need creators, creators should ignore others if they could do better on their own ($p = 1$). In other words, the fitness prospects of creators working alone can be viewed in a manner analogous to the interest yield of treasury bonds in investment decisions. This logic suggests another kind of modification of the standard discounting method. The second adaptation to the basic notion of discounting we refer to as Present Innovation Value (PIV) discounting. Let $F_{t}^{C,p}$ be the mean action fitness at period $t$ for parameter setting $\{C, p\}$. Note that $F_{t}^{1,1}$ is the fitness expectation with no interaction amongst agents. We define the PIV for any fitness curve as:
\begin{equation}
PIV(F^{C,p}) = - N + \displaystyle\sum_{t=1}^N \frac{F^{C,p}}{F^{1,1}}
\label{eq:piv}
\end{equation}
\indent Therefore, $PIV(F^{1,1}) = 0$; creators are indifferent to working alone or in a community with imitation. 

\section{Results}

All results are averages across 100 runs. 
The 3D graph and contour plot for the log$_{10}$ TTT discounting analysis of the time series for different ${C,p}$ settings are shown in Figures 1 and 2 respectively. Note that by definition a low TTT value corresponds to high mean fitness of actions across the society. The TTT method clearly demonstrates a valley in the adaptive landscape. The line running along the bottom of the valley in Figure 2 indicates, for any given value of $p$ the optimal value for $C$, and \emph{vice versa}. When $p = 1$ the optimal values of $C = 0.38$. When $C = 1$ the optimal values of $p$ is 0.19. The global optimum is at approximately $\{C,p\} = \{0.4,1.0\}$. 

\begin{figure}[ht]
\centering
\includegraphics[width=\columnwidth]{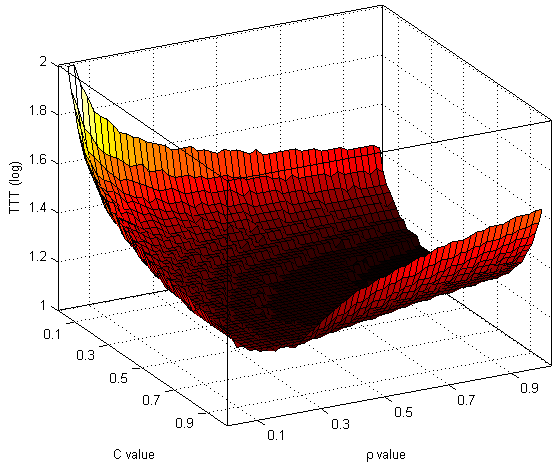}
\caption{3D graph of the log$_{10}$ Time-to-Threshold (TTT) landscape of the average mean fitness for different values of $C$ and $p$, with $\tau = 9$. The valley in the fitness landscape indicates that the optimal values of $C$ and $p$ for the society as a whole are less than their maximum values for most ${C,p}$ settings.}
\label{fig:TTT-cropped}
\end{figure}

\begin{figure}[ht]
\centering
\includegraphics[width=\columnwidth]{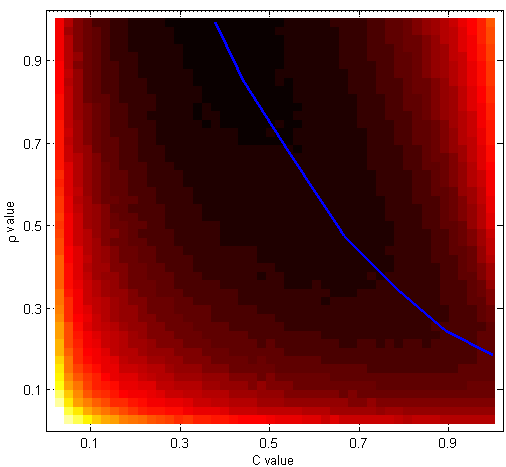}
\caption{Top-view contour plot of the log$_{10}$ Time-to-Threshold (TTT) landscape of the average mean fitness for different values of $C$ and $p$, with $\tau = 9$. The line, obtained by visually extrapolating over minimum values $C$ and $p$, indicates the set of optima.}
\label{fig:TTT-contour-line-cropped}
\end{figure}
The 3D graph and contour plot for the PIV discounting analysis of the time series for different ${C,p}$ settings are shown in Figures 3 and 4 respectively. The pattern is very  similar to that obtained with the  log$_{10}$ TTT discounting analysis. 

\begin{figure}[ht]
\centering
\includegraphics[width=\columnwidth]{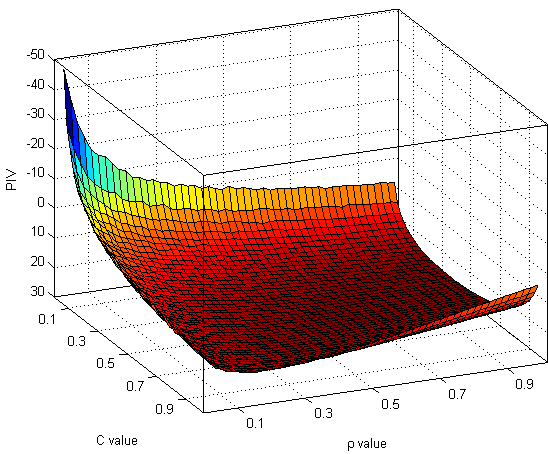}
\caption{3D graph of the Present Innovation Value (PIV) landscape of the average mean fitness for different values of $C$ and $p$. Since the $x$ axis has been inverted to aid visibility of the adaptive landscape, the valley again indicates that the optimal values of $C$ and $p$ for the society as a whole are less than their maximum values for most ${C,p}$ settings.}%
\label{fig:PIV-cropped}
\end{figure}

\begin{figure}[ht]
\centering
\includegraphics[width=\columnwidth]{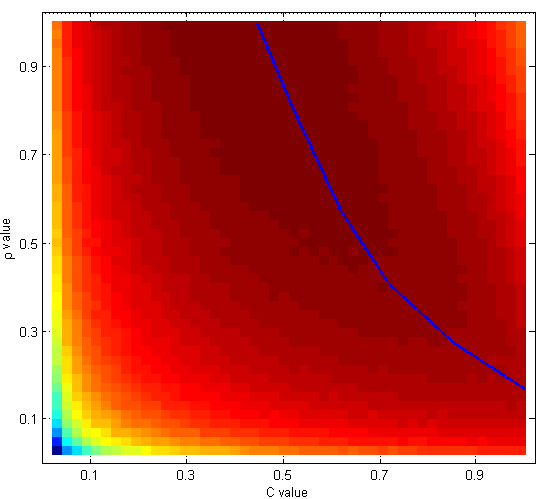}
\caption{Top-view contour plot of the Present Innovation Value (PIV) landscape of average mean fitness for different values of $C$ and $p$. The line, obtained by visually extrapolating over maximum values $C$ and $p$, indicates the set of optima.}
\label{fig:PIV-contour-line-cropped}
\end{figure}

Thus both log$_{10}$ TTT  and PIV analyses of the time series showed that, although some creativity is essential to get the fitness of cultural novelty increasing over time, more creativity is not necessarily better. For optimal mean fitness of agentsÕ actions across the society there is a tradeoff between $C$, the proportion of creators in the artificial society, and $p$, how creative these creators are.

\section{Discussion and Future Directions}

This investigation yielded results that contradict the widespread assumption that creativity is necessarily desirable. The model is highly idealized, and caution must be taken in extrapolating to human societies. The PIV results assume that creators avoid input from neighbors if doing so would maximize the fitness of their actions. In reality, creative individuals may not behave so rationally. However, the PIV results were corroborated by the TTT results, indicating that the basic pattern does not depend on the assumption of economic rationality. 

EVOC agents are too rudimentary to suffer the affective penalties of creativity but the model incorporates another important drawback to creativity mentioned in the introduction: an iteration spent inventing is an iteration not spent imitating. Creative agents, absorbed in their creative process, effectively rupture the fabric of the artificial society; they act as insulators that impede the diffusion of proven solutions. Imitators, in contrast, serve as a Òcultural memoryÓ that ensures the preservation of successful ideas. This suggests that the reason people are not more creative than they are is not just because it is difficult to be creative; there is a cost to society as well. 

Our results suggest that families, organizations, or societies may self-organize to achieve levels of both  imitation and creativity that are intermediate in order to achieve a balance between continuity and change. The results  suggest that imitation is neither just the greatest compliment, nor a form of free-riding, but a valuable social mechanism that serves innovators and imitators alike. Without invention there is nothing to imitate, but invention is considerably more effective in conjunction with imitation.

Limitation of this work include that the fitness function was static throughout a run, and agents had only one action to optimize. In real life, there are many tasks, and a division of labor such that each agent specializes in a few tasks, and imitates other agents to carry out other tasks. Another limitation is that EVOC currently does not allow an agent to imitate only certain features of an idea while retaining features the idea it is currently implementing. Creative change can break up co-adapted partial solutions. Recall that actions have a cultural version of what in biology is referred to as epistasis, wherein what is optimal with respect to one component depends on what is done with respect to another. Once both components have been optimized in a mutually beneficial way (for example, the arms are moving symmetrically), excess creativity can cause co-adapted partial solutions to break down. In future studies we will investigate the effects of using a dynamic fitness function, and enabling partial imitation. We will also compare our findings to real world data.\\
\indent If it is the case that social groups can be too creative for their own good, then expensive and widely used programs to enhance creativity through methods such as brainstorming may be counterproductive. The results of these experiments help make sense of findings that creativity is often suppressed in the classroom and in society at large, and that creative individuals often experience discrimination, or worse \cite{Craft2005,CropleyCropley2005,Scott1999,Torrance1962,Torrance1963}. (It is well-known that Einstein's dissertation was rejected by the Techniche Hochschule in Vienna; he wrote his papers on relativity while working at a patent office.) On the other hand, once the merits of ones' creative efforts become known, this individual's creativity is generally supported or even idolized. In future work we plan to investigate the hypothesis that the social practice of discouraging creativity until the creative individual has proven him- or herself serves as a form of social self-regulation ensuring that creative efforts are not squandered. Specifically, we will use EVOC to test the hypothesis that if individuals who generate creative outputs of low fitness fitness are exposed to social pressures that discourage creativity, and individuals who generate creative outputs of high fitness fitness are encouraged to be creative, the society may self-organize such that it achieves a balance of creative and uncreative individuals (such as the $C,p$ values indicated by the red line in our experiments). 

\section{Acknowledgments}
\noindent This work was supported by grants to the first author from the Social Sciences and Humanities Research Council of Canada, the Natural Sciences and Engineering Research Council of Canada, and the Flemish Fund for Scientific Research, Belgium. We thank Tiha von Ghyczy for help with the analysis.


\bibliographystyle{apacite}

\setlength{\bibleftmargin}{.125in}
\setlength{\bibindent}{-\bibleftmargin}

\bibliography{GF21}


\end{document}